\documentclass[11pt]{article}
\usepackage{graphicx}
\usepackage{amssymb}
\usepackage{amsmath}
\usepackage{graphicx}
\usepackage{amstext}
\usepackage{esint}
\usepackage[utf8]{inputenc}
\usepackage{color}
\usepackage{cite}

\renewcommand{\vec}[1]{{\bf #1}}       
\def\beq{\begin{eqnarray}}    
\def\eeq{\end{eqnarray}}      

\newcommand{\rL}{\rho_\Lambda}

\newcommand{\CC}{\Lambda}

\newcommand{\Omo}{\Omega_{m}}

\newcommand{\OLo}{\Omega_{\Lambda}}

\newcommand{\rmr}{\rho_m}

\newcommand{\rLo}{\rho_{\CC 0}}

\newcommand{\newtext}[1]{{\textcolor{black}{#1}}}
\newcommand{\newnewtext}[1]{{\textcolor{black}{#1}}}

\begin{document}



 \hyphenation{nu-cleo-syn-the-sis u-sing si-mu-la-te ma-king
cos-mo-lo-gy know-led-ge e-vi-den-ce stu-dies be-ha-vi-or
res-pec-ti-ve-ly appro-xi-ma-te-ly gra-vi-ty sca-ling
ge-ne-ra-li-zed}




\begin{center}
{\it\LARGE The $H_0$ tension in light of vacuum dynamics in the Universe} \vskip 2mm

 \vskip 8mm

\textbf{Joan Sol\`a, Adri\`a G\'omez-Valent, and Javier de Cruz P\'erez}

\vskip 0.5cm
Departament de F\'isica Qu\`antica i Astrof\'isica, and Institute of Cosmos Sciences,\\ Universitat de Barcelona, \\
Av. Diagonal 647, E-08028 Barcelona, Catalonia, Spain

\vskip0.5cm

\vskip0.4cm

E-mails:   sola@fqa.ub.edu, adriagova@fqa.ub.edu, decruz@fqa.ub.edu

 \vskip2mm

\end{center}
\vskip 15mm

\begin{quotation}
\noindent {\large\it \underline{Abstract}}.
Despite the outstanding achievements of modern cosmology, the classical dispute on the precise value of $H_0$, which is the first ever parameter of modern cosmology and one of the prime parameters in the field, still goes on and on after over half a century of measurements. Recently the dispute came to the spotlight with renewed strength owing to the significant tension (at $>3\sigma$ c.l.) between the latest Planck determination obtained from the CMB anisotropies and the local (distance ladder) measurement from the Hubble Space Telescope (HST), based on Cepheids.
In this work, we investigate the impact of the running vacuum model (RVM) and related models on such a controversy. For the RVM, the vacuum energy density $\rL$ carries a mild dependence on the cosmic expansion rate, i.e. $\rL(H)$, which allows to ameliorate the fit quality to the overall SNIa+BAO+$H(z)$+LSS+CMB cosmological observations as compared to the concordance $\CC$CDM model.
By letting the RVM to deviate from the vacuum option, the equation of state $w=-1$ continues to be favored by the overall fit.
Vacuum dynamics also predicts the following: i) the CMB range of values for $H_0$ is  more favored than the local ones, and ii) smaller values for $\sigma_8(0)$. As a result, a better account for the LSS structure formation data is achieved as compared to the $\CC$CDM, which is based on a rigid (i.e. non-dynamical) $\CC$ term.
\end{quotation}
\vskip 5mm

\newpage


\newpage


\section{Introduction}\label{intro}

The most celebrated fact of modern observational cosmology is that the universe is in accelerated expansion\,\cite{SNIaRiess,SNIaPerl}. At the same time, the most paradoxical reality check is that we do not honestly understand the primary cause for such an acceleration.  The simplest picture is to assume that it is caused by a strict cosmological term, $\CC$, in Einstein's equations, but its fundamental origin is unknown\,\cite{Weinberg2000}. Together with the assumption of the existence of dark matter (DM) and the spatial flatness of the Friedmann-Lema\^{\i}tre-Robertson-Walker (FLRW) metric (viz. the metric that expresses the homogeneity and isotropy inherent to the cosmological principle), we are led to the ``concordance'' $\CC$CDM model, i.e. the standard model of cosmology\,\cite{Peebles1993}. The model is consistent with a large body of observations, and in particular with the high precision data from the cosmic microwave background (CMB) anisotropies\,\cite{Planck2015,PlanckDE2015}.
Many alternative explanations of the cosmic acceleration beyond a  $\CC$-term are possible (including quintessence and the like, see e.g. the review\,\cite{PeeblesRatra2003}) and are called dark energy (DE)\,\cite{DEBook}.

The current situation with cosmology is reminiscent of the prediction by the famous astronomer A. Sandage in the sixties, who asserted that the main task of future observational cosmology would be the search for two parameters: the Hubble constant $H_0$ and the deceleration parameter $q_0$\,\cite{Sandage1961}. The first of them is the most important distance (and time) scale in cosmology prior to any other cosmological quantity. Sandage's last published value with Tammann (in 2010) is $62.3$ km/s/Mpc\,\cite{TammannSandage2010} -- slightly revised in Ref.\,\cite{TammannReindl2013} as $H_0 = 64.1 \pm
2.4$ km/s/Mpc. There is currently a significant tension between CMB measurements of $H_0$\,\cite{Planck2015,Planck2016} -- not far away from this value -- and local determinations emphasizing a higher range above $70$ km/s/Mpc\,\cite{RiessH02016,RiessH02011}. As for $q_0$, its measurement is tantamount to determining $\CC$ in the context of the concordance model.
On fundamental grounds, however, understanding the value of $\CC$ is not just a matter of observation; in truth and in fact, it embodies one of the most important and unsolved conundrums of theoretical physics and cosmology: the cosmological constant problem, see e.g.\,\cite{Weinberg2000,PeeblesRatra2003,Padmanabhan2003,JSPRev2013}. The problem is connected to the fact that the $\CC$-term is usually associated with the vacuum energy density, $\rL=\CC/(8\pi G)$, with $G$ Newton's coupling. The prediction for  $\rL$ in quantum field theory (QFT) overshoots the measured value $\rL\sim 10^{-47}$ GeV$^4$ (in natural units $c=\hbar=1$) by many orders of magnitude\,\cite{JSPRev2013}.

Concerning the prime parameter $H_0$, the tension among the different measurements is inherent to its long and tortuous history. Let us only recall that after Baade's revision (by a factor of one half\,\cite{Baade1944}) of the exceedingly large value $\sim 500$ km/s/Mpc originally estimated by Hubble (which implied a universe of barely two billion years old only), the Hubble parameter was subsequently lowered to $75$ km/s/Mpc and finally to $H_0=55\pm 5$ km/s/Mpc, where it remained for 20 years (until 1995), mainly under the influence of Sandage's devoted observations\,\cite{Tammann1996}. Shortly after that period the first measurements of the nonvanishing, positive, value of $\CC$ appeared\,\cite{SNIaRiess,SNIaPerl} and the typical range for $H_0$ moved upwards to $\sim 65$ km/s/Mpc. In the meantime, many different observational values of $H_0$ have piled up in the literature using different methods (see e.g. the median statistical analysis of $>550$ measurements considered in \cite{ChenRatra2011,BethapudiDesai2017}). As mentioned above, two kinds of \emph{precision} (few percent level) measurements of $H_0$ have generated considerable perplexity in the recent literature, specifically between the latest Planck values ($H^{\rm Planck}_0$) obtained from the CMB anisotropies, and the local HST measurement (based on distance ladder estimates from Cepheids). The latter, obtained by Riess {\it et al.}\,\cite{RiessH02016}, is $H_0 = 73.24\pm 1.74$\, km/s/Mpc and will be denoted $H_0^{\rm Riess}$. It can be compared with the CMB value  $H_0 = 67.51\pm 0.64$ km/s/Mpc, as extracted from Planck 2015 TT,TE,EE+lowP+lensing data\,\cite{Planck2015}, or with
$H_0 = 66.93 \pm 0.62$ km/s/Mpc, based on Planck 2015 TT,TE,EE+SIMlow data\,\cite{Planck2016}. In both cases there is a tension above $3\sigma$ c.l. (viz. $3.1\sigma$ and $3.4\sigma$, respectively) with respect to the local measurement. This situation, and in general a certain level of tension with some independent observations
in intermediate cosmological scales, has stimulated a number of discussions and possible solutions in the literature, see e.g.\,\cite{Melchiorri2016,Bernal2016,Shafieloo2017,Cardona2017,Melchiorri2017a,Melchiorri2017b,Zhang2017,Wang2017,Feeney2017}.

We wish to reexamine here the $H^{\rm Riess}_0-H^{\rm Planck}_0$ tension, but not as an isolated conflict between two particular sources of observations, but rather in light of the overall fit to the cosmological data SNIa+BAO+$H(z)$+LSS+CMB. Recently, it has been demonstrated that by letting the cosmological vacuum energy density to slowly evolve with the expansion rate, $\rL=\rL(H)$, the global fit can be improved with respect to the $\CC$CDM at a confidence level of $3-4\sigma$ \,\cite{ApJL2015,ApJ2017,MPLA2017,JSPRev2016,PRD2017}.
We devote this work to show that the dynamical vacuum models (DVMs) can still give a better fit to the overall data, even if the local HST measurement of the Hubble parameter is taken into account. {However we find that our best-fit values of $H_0$ are much closer to the value extracted from CMB measurements \cite{Planck2015,Planck2016}}. Our analysis also corroborates that the large scale structure formation data (LSS) are crucial in distinguishing the rigid vacuum option from the dynamical one.


\begin{table*}
\begin{center}
\begin{scriptsize}
\resizebox{1\textwidth}{!}{
\begin{tabular}{ |c|c|c|c|c|c|c|c|c|c|}
\multicolumn{1}{c}{Model} &  \multicolumn{1}{c}{$H_0$(km/s/Mpc)} &  \multicolumn{1}{c}{$\omega_b$} & \multicolumn{1}{c}{{\small$n_s$}}  &  \multicolumn{1}{c}{$\Omega_m^0$} &\multicolumn{1}{c}{$\nu_i$} &\multicolumn{1}{c}{$w$} &\multicolumn{1}{c}{$\chi^2_{\rm min}/dof$} & \multicolumn{1}{c}{$\Delta{\rm AIC}$} & \multicolumn{1}{c}{$\Delta{\rm BIC}$}\vspace{0.5mm}
\\\hline
$\Lambda$CDM  & $68.83\pm 0.34$ & $0.02243\pm 0.00013$ &$0.973\pm 0.004$& $0.298\pm 0.004$ & - & -1  & 84.40/85 & - & - \\
\hline
XCDM  & $67.16\pm 0.67$& $0.02251\pm0.00013 $&$0.975\pm0.004$& $0.311\pm0.006$ & - &$-0.936\pm{0.023}$  & 76.80/84 & 5.35 & 3.11 \\
\hline
RVM  & $67.45\pm 0.48$& $0.02224\pm0.00014 $&$0.964\pm0.004$& $0.304\pm0.005$ &$0.00158\pm 0.00041 $ & -1  & 68.67/84 & 13.48 & 11.24 \\
\hline
$Q_{dm}$  & $67.53\pm 0.47$& $0.02222\pm0.00014 $&$0.964\pm0.004$& $0.304\pm0.005$ &$0.00218\pm 0.00058 $&-1  & 69.13/84 & 13.02 &10.78 \\
\hline
$Q_\Lambda$  & $68.84\pm 0.34$& $0.02220\pm0.00015 $&$0.964\pm0.005$& $0.299\pm0.004$ &$0.00673\pm 0.00236 $& -1  &  76.30/84 & 5.85 & 3.61\\
\hline
$w$RVM  & $67.08\pm 0.69$& $0.02228\pm0.00016 $&$0.966\pm0.005$& $0.307\pm0.007$ &$0.00140\pm 0.00048 $ & $-0.979\pm0.028$ & 68.15/83 & 11.70 & 7.27 \\
\hline
$w{Q_{dm}}$  & $67.04\pm 0.69$& $0.02228\pm0.00016 $&$0.966\pm0.005$& $0.308\pm0.007$ &$0.00189\pm 0.00066 $& $-0.973\pm 0.027$ & 68.22/83 & 11.63 & 7.20\\
\hline
$w{Q_\Lambda}$  & $67.11\pm 0.68$& $0.02227\pm0.00016 $&$0.965\pm0.005$& $0.313\pm0.006$ &$0.00708\pm 0.00241 $& $-0.933\pm0.022$ &   68.24/83 & 11.61 & 7.18\\
\hline
\end{tabular}}
\caption{{\scriptsize Best-fit values for the $\CC$CDM, XCDM, the three dynamical vacuum models (DVMs) and the three dynamical quasi-vacuum models ($w$DVMs), including their statistical significance ($\chi^2$-test and Akaike and Bayesian information criteria AIC and BIC).
For detailed description of the data and a full list of references, see \cite{ApJ2017} and \cite{PRD2017}. The quoted number of degrees of freedom ($dof$) is equal to the number of data points minus the number of independent fitting parameters ($4$ for the $\CC$CDM, 5 for the XCDM and the DVMs, and 6 for the $w$DVMs). For the CMB data we have used the marginalized mean values and {covariance matrix} for the parameters of the compressed likelihood for Planck 2015 TT,TE,EE + lowP+ lensing data from\,\cite{WangDai2016}. Each best-fit value and the associated uncertainties have been obtained by marginalizing over the remaining parameters.}}
\end{scriptsize}
\end{center}
\label{tableFit1}
\end{table*}


\section{Dynamical vacuum models and beyond}
Let us consider a generic cosmological framework described by the spatially flat FLRW metric, in which matter is exchanging energy with a dynamical DE medium with a phenomenological equation of state (EoS) $p_{\CC}=w\rho_{\CC}$, where $w=-1+\epsilon$ (with $|\epsilon|\ll1$). Such medium is therefore of quasi-vacuum type, and for $w=-1$ (i.e. $\epsilon=0$) we precisely recover the genuine vacuum case. Owing, however, to the exchange of energy with matter, $\rL=\rL(\zeta)$ is in all cases a {\it dynamical} function that depends on a cosmic variable $\zeta=\zeta(t)$.  We will identify the nature of $\zeta(t)$ later on, but its presence clearly indicates that $\rL$ is no longer associated to a strictly rigid cosmological constant as in the $\CC$CDM. The Friedmann and acceleration equations read, however, formally identical to the standard case:
\begin{eqnarray}
&&3H^2=8\pi\,G\,(\rho_m+\rho_r+\rho_\Lambda(\zeta))\label{eq:FriedmannEq}\\
&&3H^2+2\dot{H}=-8\pi\,G\,(p_r + p_\Lambda(\zeta))\,.\label{eq:PressureEq}
\end{eqnarray}
Here $H=\dot{a}/a$ is the Hubble function, $a(t)$ the scale factor as a function of the cosmic time, $\rho_r$ is the energy density of the radiation component (with pressure $p_r=\rho_r/3$), and $\rho_m=\rho_b+\rho_{dm}$ involves the contributions from baryons and cold DM. The local conservation law associated to the above equations reads:
\begin{equation}\label{eq:GeneralCL}
\dot{\rho}_r + 4H\rho_r + \dot{\rho}_m + 3H\rho_m = Q\,,
\end{equation}
where
\begin{equation}\label{eq:Source}
Q = -\dot{\rho}_\CC - 3H(1+w)\rho_\CC\,.
\end{equation}
For $w=-1$ the last equation boils down to just $Q= -\dot{\rho}_\CC$, which is nonvanishing on account of $\rL(t)=\rL(\zeta(t))$.
\begin{table*}
\begin{center}
\begin{scriptsize}
\resizebox{1\textwidth}{!}{
\begin{tabular}{ |c|c|c|c|c|c|c|c|c|c|}
\multicolumn{1}{c}{Model} &  \multicolumn{1}{c}{$H_0$(km/s/Mpc)} &  \multicolumn{1}{c}{$\omega_b$} & \multicolumn{1}{c}{{\small$n_s$}}  &  \multicolumn{1}{c}{$\Omega_m^0$} &\multicolumn{1}{c}{$\nu_i$} &\multicolumn{1}{c}{$w$} &\multicolumn{1}{c}{$\chi^2_{\rm min}/dof$} & \multicolumn{1}{c}{$\Delta{\rm AIC}$} & \multicolumn{1}{c}{$\Delta{\rm BIC}$}\vspace{0.5mm}
\\\hline
$\Lambda$CDM  & $68.99\pm 0.33$ & $0.02247\pm 0.00013$ &$0.974\pm 0.003$& $0.296\pm 0.004$ & - & -1  & 90.59/86 & - & - \\
\hline
XCDM  & $67.98\pm 0.64$& $0.02252\pm0.00013 $&$0.975\pm0.004$& $0.304\pm0.006$ & - &$-0.960\pm{0.023}$  & 87.38/85 & 0.97 & -1.29 \\
\hline
RVM  & $67.86\pm 0.47$& $0.02232\pm0.00014 $&$0.967\pm0.004$& $0.300\pm0.004$ &$0.00133\pm 0.00040 $ & -1  & 78.96/85 & 9.39 & 7.13 \\
\hline
$Q_{dm}$  & $67.92\pm 0.46$& $0.02230\pm0.00014 $&$0.966\pm0.004$& $0.300\pm0.004$ &$0.00185\pm 0.00057 $&-1  & 79.17/85 & 9.18 & 6.92 \\
\hline
$Q_\Lambda$  & $69.00\pm 0.34$& ${0.02224}\pm0.00016 $&$0.965\pm0.005$& $0.297\pm0.004$ &$0.00669\pm 0.00234 $& -1  &  82.48/85 & 5.87 & 3.61\\
\hline
$w$RVM  & $67.95\pm 0.66$& $0.02230\pm0.00015 $&$0.966\pm0.005$& $0.300\pm0.006$ &$0.00138\pm 0.00048 $ & $-1.005\pm0.028$ & 78.93/84 & 7.11 & 2.66 \\
\hline
$w{Q_{dm}}$  & $67.90\pm 0.66$& $0.02230\pm0.00016 $&$0.966\pm0.005$& $0.300\pm0.006$ &$0.00184\pm 0.00066 $& $-0.999\pm 0.028$ & 79.17/84 & 6.88 & 2.42\\
\hline
$w{Q_\Lambda}$  & $67.94\pm 0.65$& $0.02227\pm0.00016 $&$0.966\pm0.005$& $0.306\pm0.006$ &$0.00689\pm 0.00237 $& $-0.958\pm0.022$ &   78.98/84 & 7.07 & 2.61\\
\hline
\end{tabular}}

\caption{{\scriptsize The same as Table 1 but adding the $H_0^{\rm Riess}$ local measurement from Riess {\it et al.}\, \cite{RiessH02016}.}}
\end{scriptsize}
\end{center}
\label{tableFit2}
\end{table*}
The simplest case is, of course, that of the concordance model, in which $\rL=\rLo=$const and $w=-1$, so that $Q=0$ trivially. However, for $w\neq -1$ we can also have $Q=0$ in a nontrivial situation, which follows from solving Eq.\,(\ref{eq:Source}). It corresponds to the XCDM parametrization\,\cite{XCDM}, in which the DE density is dynamical and self-conserved. It is easily found in terms of the scale factor:
\begin{equation}\label{eq:rhoXCDM}
\rL^{XCDM}(a)=\rLo\,a^{-3(1+w)}=\rLo\,a^{-3\epsilon}\,,
\end{equation}
where $\rLo$ is the current value.
From (\ref{eq:GeneralCL}) it then follows that the total matter component is also conserved. After equality it leads to separate conservation of cold matter and radiation.   In general, $Q$ can be a nonvanishing interaction source allowing energy exchange between matter and the quasi-vacuum medium under consideration; $Q$ can either be given by hand (e.g. through an {\it ad hoc} ansatz), or can be suggested by some specific theoretical framework. In any case the interaction source must satisfy $0<|Q|\ll\dot{\rho}_m$ since we do not wish to depart too much from the concordance model. Despite matter is exchanging energy with the vacuum or quasi-vacuum medium, we shall assume that radiation and baryons are separately self-conserved, i.e. $\dot{\rho}_r + 4H\rho_r =0$ and $\dot{\rho}_b + 3H\rho_b =0$, so that their energy densities evolve in the standard way: $\rho_r(a)=\rho_{r0}\,a^{-4}$ and $\rho_b(a) = \rho_{b0}\,a^{-3}$. The dynamics of $\rL$ can therefore be  associated to the exchange of energy exclusively with the DM (through the nonvanishing source $Q$) and/or with the possibility that the DE medium is not exactly the vacuum, $w\neq -1$, but close to it $|\epsilon|\ll 1$. Under these conditions, the coupled system of conservation equations (\ref{eq:GeneralCL})-(\ref{eq:Source}) reduces to
\begin{eqnarray}
&&\dot{\rho}_{dm}+3H\rho_{dm}=Q\label{eq:Qequations1}\\
&&\dot{\rho}_\CC + 3H\epsilon\rho_\CC=-Q\,.\label{eq:Qequations2}
\end{eqnarray}
\begin{figure*}
\begin{center}
\label{FigLSS1}
\includegraphics[width=5.7in]{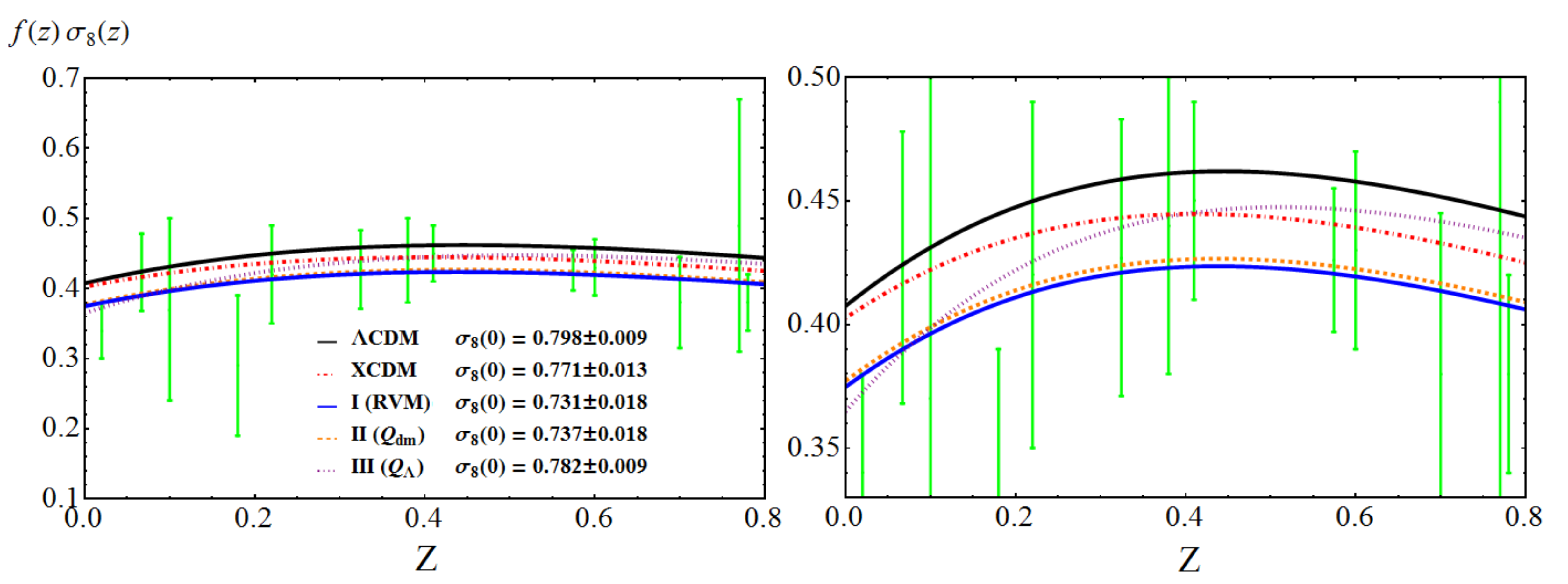}\ \ 
\caption{\scriptsize {\bf Left}: The LSS  structure formation data ($f(z)\sigma_8(z)$) versus the predicted curves by Models I, II and III, see equations (\ref{eq:QforModelRVM})-(\ref{eq:QforModelQL}) for the case $w=-1$, i.e. the dynamical vacuum models (DVMs), using the best-fit values in Table 1. The XCDM curve is also shown. The values of $\sigma_8(0)$ that we obtain for the models are also indicated. {\bf Right}: Zoomed window of the plot on the left, which allows to better distinguish the various models.}
\end{center}
\end{figure*}

In the following we shall for definiteness focus our study of the dynamical vacuum (and quasi-vacuum) models to the three interactive sources:
\begin{eqnarray}\label{eq:QforModelRVM}
&&{\rm Model\ I\ \ }(w{\rm RVM}):\ Q=\nu\,H(3\rho_{m}+4\rho_r)\label{eq:QforModelQdm}\\
&&{\rm Model\ II\ \ }(wQ_{dm}):\ Q_{dm}=3\nu_{dm}H\rho_{dm}\\
&&{\rm Model\ III\ \ }(wQ_{\CC}):\ Q_{\CC}=3\nu_{\CC}H\rho_{\CC}\,.\label{eq:QforModelQL}
\end{eqnarray}
Here $\nu_i=\nu, \nu_{dm},\nu_{\CC}$ are small dimensionless constants, $|\nu_i|\ll1$, which are determined from the overall fit to the data, see e.g. Tables 1 and 2.
The ordinal number names I, II and III will be used for short, but the three model names are preceded by $w$  to recall that, in the general case, the equation of state (EoS) is near the vacuum one (that is, $w=-1+\epsilon$). These dynamical quasi-vacuum models  are also denoted as $w$DVMs. In the particular case $w=-1$ (i.e. $\epsilon=0$) we recover the dynamical vacuum models (DVMs), which were previously studied in detail in\,\cite{PRD2017}, and in this case the names of the models will not be preceded by $w$.

 In all of the above ($w$)DVMs, the cosmic variable $\zeta$ can be taken to be the scale factor, $\zeta=a(t)$, since they are all analytically solvable in terms of it, as we shall see in a moment. Model I with $w=-1$ is the running vacuum model (RVM), see\,\cite{JSPRev2013,SolGomRev2015,PRD2017,JSPRev2016}. It is special in that the interaction source indicated in (\ref{eq:QforModelRVM}) is not {\it ad hoc} but follows from an expression for the dynamical vacuum energy density, $\rL(\zeta)$, in which
$\zeta$ is not just the scale factor but the full Hubble rate: $\zeta=H(a)$.  The explicit RVM form reads
\begin{equation}\label{eq:RVMvacuumdadensity}
\rho_\CC(H) = \frac{3}{8\pi{G}}\left(c_{0} + \nu{H^2}\right)\,.
\end{equation}
\newline
The additive constant $c_0=H_0^2\left(\OLo^0-\nu\right)$ is fixed from the condition $\rL(H_0)=\rLo$, with $\Omega^{0}_\Lambda=1-\Omega^{0}_m-\Omega^{0}_r$. Combining the Friedmann and acceleration equations (\ref{eq:FriedmannEq})-(\ref{eq:PressureEq}), we find $\dot{H}=-(4\pi G/3)$
$\left(3\rho_m+4\rho_r+3\epsilon\rho_\CC\right)$,
and upon differentiating (\ref{eq:RVMvacuumdadensity}) with respect to the cosmic time we are led to
$\dot{\rho}_\CC=-\nu\,H\left(3\rho_m+4\rho_r+3\epsilon\rho_\CC\right)$.
Thus, for $\epsilon=0$ (vacuum case) we indeed find $\dot{\rho}_\CC=-Q$ for $Q$ as in (\ref{eq:QforModelRVM}). However, for the quasi-vacuum case ($0<|\epsilon|\ll1$) Eq.\,(\ref{eq:Qequations2}) does not hold if $\rL(H)$ adopts the form (\ref{eq:RVMvacuumdadensity}). This RVM form is in fact specific to the pure vacuum EoS ($w=-1$), and it can be motivated in QFT in curved spacetime through a renormalization group equation for $\rL(H)$, what explains the RVM name\,\cite{JSPRev2013}. In it, $\nu$ plays the role of the $\beta$-function coefficient for the running of $\rL$ with the Hubble rate. Thus, we naturally expect $|\nu|\ll1$ in QFT, see \cite{JSPRev2013,Fossil07}. Interestingly, the
RVM  form (\ref{eq:RVMvacuumdadensity}) can actually be extended with higher powers of $H^n$ (typically $n=4$) to provide an effective description of the cosmic evolution from the inflationary universe up to our days\,\cite{LimBasSol2013,SolGomRev2015}. Models II and III are purely phenomenological models instead, in which the interaction source $Q$ is introduced by hand, see e.g. Refs.\,\cite{Salvatelli2014,Murgia2016,Li2016,Melchiorri2017b} and references therein.

The energy densities for the $w$DVMs can be computed straightforwardly. For simplicity, we shall quote here the leading parts only. The exact formulas containing the radiation terms are more cumbersome. In the numerical analysis we have included the full expressions. Details will be shown elsewhere. For the matter densities, we find:
\begin{eqnarray}
\rho^{\rm I}_{dm}(a) &=& \rho_{dm0}\,a^{-3(1-\nu)} + \rho_{b0}\left(a^{-3(1-\nu)} - a^{-3}\right) \nonumber \\
\rho^{\rm II}_{dm}(a) &=& \rho_{dm0}\,a^{-3(1-\nu_{dm})}
\label{eq:rhoms} \\
\rho^{\rm III}_{dm}(a) &=&\rho_{dm0}\,a^{-3} + \frac{\nu_\CC}{\nu_\CC + w}\rLo\left(a^{-3}-a^{-3(\epsilon + \nu_\CC)}\right)\nonumber\,,
\end{eqnarray}
and for the quasi-vacuum energy densities:
\begin{eqnarray}\label{eq:rhowLdensities}
\rho^{\rm I}_\CC(a) &=& \rLo{a^{-3\epsilon}} - \frac{\nu\,\rho_{m0}}{\nu + w}\left(a^{-3(1-\nu)}- a^{-3\epsilon}\right) \nonumber\\
\rho^{\rm II}_\CC(a)&=& \rLo{a^{-3\epsilon}} - \frac{\nu_{dm}\,\rho_{dm0}}{\nu_{dm} + w}\,\left(a^{-3(1-\nu_{dm})}- a^{-3\epsilon}\right)\\
\rho^{\rm III}_\CC(a) &=&\rLo\,{a^{-3(\epsilon + \nu_\CC)}}\,.\nonumber
\end{eqnarray}
Two specific dimensionless  parameters enter each formula, $\nu_{i}=(\nu,\nu_{dm},\nu_\CC)$  and $w=-1+\epsilon$. They are part of the fitting vector of free parameters for each model, as explained in detail in the caption of Table 1. For $\nu_{i}\to 0$ the models become noninteractive and they all reduce to the XCDM model case (\ref{eq:rhoXCDM}). For $w=-1$ we recover the DVMs results previously studied in \cite{PRD2017}.  Let us also note that for $\nu_{i}>0$ the vacuum decays into DM (which is thermodynamically favorable\,\cite{PRD2017}) whereas for $\nu_{i}<0$ is the other way around. Furthermore, when $w$ enters the fit, the effective behavior of the $w$DVMs is quintessence-like for $w>-1$ (i.e. $\epsilon>0$) and phantom-like for $w<-1$ ($\epsilon<0$).

Given the energy densities (\ref{eq:rhoms}) and (\ref{eq:rhowLdensities}), the Hubble function immediately follows. For example, for Model I:
\begin{equation}\label{eq:HubbewRVM}
H^2(a) = H_0^2\left[a^{-3\epsilon} + \frac{w}{w+\nu}\Omega^{0}_m\left(a^{-3(1-\nu)}- a^{-3\epsilon}\right)\right]\,.
\end{equation}
Similar formulas can be obtained for Models II and III. For $w=-1$ they all reduce to the DVM forms previously found in \cite{PRD2017}. And of course they all ultimately boil down to the $\CC$CDM form in the limit $(w, \nu_i)\to (-1, 0)$.



\begin{table*}
\begin{center}
\begin{scriptsize}
\resizebox{1\textwidth}{!}{
\begin{tabular}{ |c|c|c|c|c|c|c|c|c|c|}
\multicolumn{1}{c}{Model} &  \multicolumn{1}{c}{$H_0$(km/s/Mpc)} &  \multicolumn{1}{c}{$\omega_b$} & \multicolumn{1}{c}{{\small$n_s$}}  &  \multicolumn{1}{c}{$\Omega_m^0$} &\multicolumn{1}{c}{$\nu_i$} &\multicolumn{1}{c}{$w$} &\multicolumn{1}{c}{$\chi^2_{\rm min}/dof$} & \multicolumn{1}{c}{$\Delta{\rm AIC}$} & \multicolumn{1}{c}{$\Delta{\rm BIC}$}\vspace{0.5mm}
\\\hline
$\Lambda$CDM  & $68.23\pm 0.38$ & $0.02234\pm 0.00013$ &$0.968\pm 0.004$& $0.306\pm 0.005$ & - & -1  & 13.85/11 & - & - \\
\hline
RVM  & $67.70\pm 0.69$& $0.02227\pm0.00016 $&$0.965\pm0.005$& $0.306\pm0.005$ &$0.0010\pm 0.0010 $ & -1  & 13.02/10 & -3.84 & -1.88 \\
\hline
$Q_\Lambda$  & $68.34\pm 0.40$& $0.02226\pm0.00016 $&$0.965\pm0.005$& $0.305\pm0.005$ &$0.0030\pm 0.0030 $& -1  &  12.91/10 & -3.73 & -1.77 \\
\hline
$w$RVM  & $66.34\pm 2.30$& $0.02228\pm0.00016 $&$0.966\pm0.005$& $0.313\pm0.012$ &$0.0017\pm 0.0016 $ & $-0.956\pm0.071$ & 12.65/9 & -9.30 & -4.22 \\
\hline
$w{Q_\Lambda}$  & $66.71\pm 1.77$& $0.02226\pm0.00016 $&$0.965\pm0.005$& $0.317\pm0.014$ &$0.0070\pm 0.0054 $& $-0.921\pm0.082$ &   12.06/9 & -8.71 & -3.63\\
\hline
$\Lambda$CDM* & $68.46\pm 0.37$ & $0.02239\pm 0.00013$ &$0.969\pm 0.004$& $0.303\pm 0.005$ & - & -1 & 21.76/12 & - & - \\
\hline
RVM*  & $68.48\pm 0.67$& $0.02240\pm0.00015 $&$0.969\pm0.005$& $0.303\pm0.005$ &$0.0000\pm 0.0010 $ & -1  & 21.76/11 & -4.36 & -2.77 \\
\hline
$Q_\Lambda$*  & $68.34\pm 0.39$& $0.02224\pm0.00016 $&$0.966\pm0.005$& $0.302\pm0.005$ &$0.0034\pm 0.0030 $& -1  &  20.45/11 & -3.05 & -1.46 \\
\hline
Ia ($w$RVM*)  & $70.95\pm 1.46$& $0.02231\pm0.00016 $&$0.967\pm0.005$& $0.290\pm0.008$ &$-0.0008\pm 0.0010 $ & $-1.094\pm0.050$ & 18.03/10 & -5.97 & -1.82 \\
\hline
IIIa ($w{Q_\Lambda}$*)  & $70.27\pm 1.33$& $0.02228\pm0.00016 $&$0.966\pm0.005$& $0.291\pm0.010$ &$-0.0006\pm 0.0042 $& $-1.086\pm0.065$ &   18.64/10 & -6.58  & -2.43 \\
\hline
\end{tabular}}

\caption{{\scriptsize Best-fit values for the $\CC$CDM and models RVM, Q$_{\Lambda}$, $w$RVM and $w$Q$_\Lambda$  by making use of the CMB+BAO data only. In contrast to Tables 1-2, we now fully dispense with the LSS data (see \cite{ApJ2017,PRD2017}) to test its effect. The starred/non-starred cases correspond respectively to adding or not the local value $H_0^{\rm Riess}$ from \cite{RiessH02016} as data point in the fit. The AIC and BIC differences of the starred models are computed with respect to the $\Lambda$CDM*. We can see that under these conditions models tend to have $\Delta$AIC, $\Delta$BIC$<0$, including the last two starred scenarios, which are capable of significantly approaching $H_0^{\rm Riess}$}.}
\end{scriptsize}
\end{center}
\label{tableFit3}
\end{table*}

\section{Structure formation: the role of the LSS data}
The analysis of structure formation plays a crucial role in comparing the various models. For the $\CC$CDM and XCDM  we use the standard perturbations equation\,\cite{Peebles1993}
\begin{equation}\label{diffeqLCDM}
\ddot{\delta}_m+2H\,\dot{\delta}_m-4\pi
G\rmr\,\delta_m=0\,,
\end{equation}
with, however, the Hubble function corresponding to each one of these models.
For the $w$DVMs, a step further is needed: the perturbations equation not only involves the modified Hubble function but the equation itself becomes modified. Trading the cosmic time for the scale factor and extending the analysis of \cite{PRD2017,GomSolBas2015,GomSol2015} for the case  $w\neq -1$ ($\epsilon\neq 0$), we find
\begin{equation}\label{diffeqD}
\delta^{\prime\prime}_m + \frac{A(a)}{a}\delta_{m}^{\prime} + \frac{B(a)}{a^2}\delta_m = 0  \,,
\end{equation}
where the prime denotes differentiation with respect to the scale factor,
and the functions $A(a)$ and $B(a)$ are found to be as follows:
\begin{eqnarray}
&&A(a) = 3 + \frac{aH^{'}}{H} + \frac{\Psi}{H} - 3r\epsilon\label{eq:Afunction}\\
&&B(a) = -\frac{4\pi{G}\rho_m }{H^2} + 2\frac{\Psi}{H} + \frac{a\Psi^{'}}{H} -15r\epsilon - 9\epsilon^{2}r^{2} +3\epsilon(1+r)\frac{\Psi}{H} -3r\epsilon\frac{aH^{'}}{H}\,. \label{eq:Bfunction}
\end{eqnarray}
Here $r \equiv \rho_\Lambda/\rho_m$ and  $\Psi\equiv -\dot{\rho}_{\Lambda}/{\rmr}$. {For $\nu_i=0$ we have $\Psi=3Hr\epsilon$}, and after a straightforward calculation one can show that (\ref{diffeqD})  can be brought to the standard form  Eq.\,(\ref{diffeqLCDM}).


\begin{figure}
\begin{center}
\label{FigLSS2}
\includegraphics[width=3.3in]{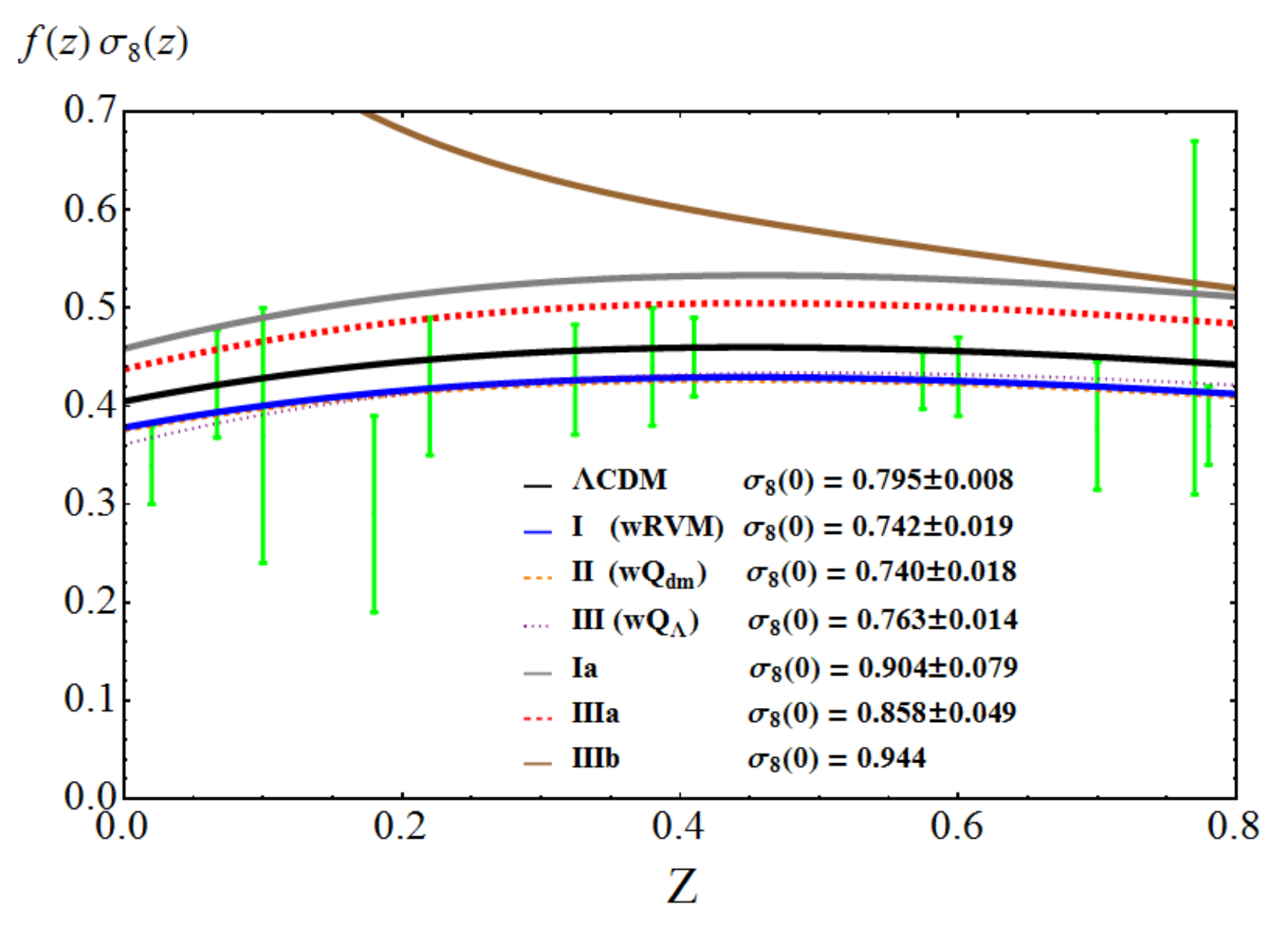}
\caption{\scriptsize The LSS  structure formation data ($f(z)\sigma_8(z)$) and the theoretical predictions for models (\ref{eq:QforModelRVM})-(\ref{eq:QforModelQL}), using the best-fit values in Tables 2 and 3. The curves for the cases Ia, IIIa correspond to special scenarios for Models I and III where the agreement with the  Riess {\it et al.} local value $H_0^{\rm Riess}$\,\cite{RiessH02016} is better (cf. Table 3). The price, however, is that the concordance with the LSS data is now spoiled. Case IIIb is our theoretical prediction for the scenario proposed in \cite{Melchiorri2017b}, aimed at optimally relaxing the tension with $H_0^{\rm Riess}$. Unfortunately, the last three scenarios lead to phantom-like DE and are in serious disagreement with the LSS data.
}
\end{center}
\end{figure}


To solve the above perturbations equations we have to fix the initial conditions on $\delta_m$ and ${\delta}'_m$ for each model at high redshift, namely when non-relativistic matter dominates over radiation and DE, see\,\cite{PRD2017}.
Functions (\ref{eq:Afunction}) and (\ref{eq:Bfunction}) are then approximately constant and Eq.\,(\ref{diffeqD}) admits power-law solutions $\delta_m(a) = a^{s}$. From explicit calculation we find that the values of $s$ for each model turn out to be:
\begin{eqnarray}\label{eq:svalues}
s^{\rm I} &=& 1 + \frac{3}{5}\nu\left(\frac{1}{w} -4\right) + \mathcal{O}(\nu^2)\nonumber  \\
s^{\rm II} &=& 1 -\frac{3}{5}\nu_{dm}\left(  1 + 3\frac{\Omega^{0}_{dm}}{\Omega^{0}_{m}} -\frac{1}{w}   \right) + \mathcal{O}({\nu_{dm}}^{2}) \\
s^{\rm III} &=&1\nonumber\,.
\end{eqnarray}
We can check that for $w=-1$ all of the above equations (\ref{diffeqD})-(\ref{eq:svalues}) render the DVM results previously found in \cite{PRD2017}. The generalization that we have made to $w\neq -1$ ($\epsilon\neq 0$) has introduced several nontrivial extra terms in equations (\ref{eq:Afunction})-(\ref{eq:svalues}).

The analysis of the linear LSS regime is usually implemented with the help of the weighted linear growth $f(z)\sigma_8(z)$, where $f(z)=d\ln{\delta_m}/d\ln{a}$ is the growth factor and $\sigma_8(z)$ is the rms mass fluctuation on $R_8=8\,h^{-1}$ Mpc scales. It is computed as follows (see e.g. \cite{PRD2017,ApJ2017}):
\begin{equation}
\begin{small}\sigma_{\rm 8}(z)=\sigma_{8, \CC}
\frac{\delta_m(z)}{\delta^{\CC}_{m}(0)}
\sqrt{\frac{\int_{0}^{\infty} k^{n_s+2} T^{2}(\vec{p},k)
W^2(kR_{8}) dk} {\int_{0}^{\infty} k^{n_{s,\CC}+2} T^{2}(\vec{p}_\Lambda,k) W^2(kR_{8,\Lambda}) dk}}\,,\label{s88general}
\end{small}\end{equation}
where $W$ is a top-hat smoothing function and $T(\vec{p},k)$ the transfer function. The fitting parameters for each model are contained in $\vec{p}$.
Following the mentioned references, we have defined as fiducial model the $\CC$CDM at fixed parameter values from the Planck 2015 TT,TE,EE+lowP+lensing data\,\cite{Planck2015}. These fiducial values are collected in  $\vec{p}_\CC$.
In Figs. 1-2 we display  $f(z)\sigma_8(z)$ for the various models using the fitted values of Tables 1-3. We remark that our BAO and LSS data include the bispectrum data points from Ref.\,\cite{GilMarin2017MNRAS}
--see \cite{PRD2017} for a full-fledged explanation of our data sets.  In the next section, we discuss our results for the various models and assess their ability to improve the $\CC$CDM fit as well as their impact on the $H_0$ tension.


\begin{figure}
\begin{center}
\label{contours}
\includegraphics[width=5in]{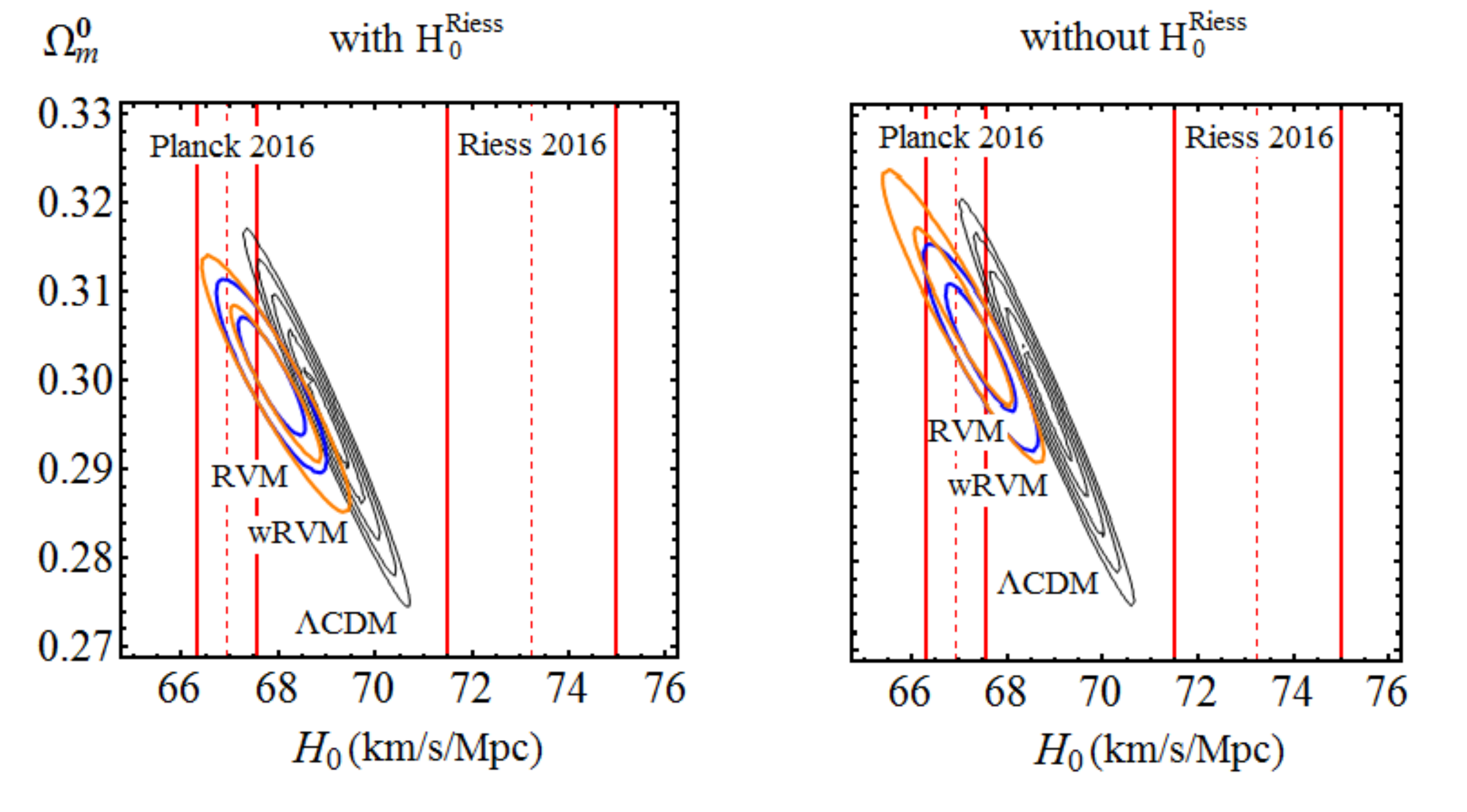}
\caption{\scriptsize Contour plots for the RVM (blue) and $w$RVM (orange) up to $2\sigma$, and for the $\CC$CDM (black) up to $5\sigma$ in the $(H_0,\Omo^0)$-plane. {Shown are the two relevant cases under study: the plot on the left is for when the local $H_0$ value of Riess {\it et al.}\,\cite{RiessH02016} is included in the fit (cf. Table 2), and the plot on the right is for when that local value is {\it not} included (cf. Table 1). Any attempt at reaching the $H_0^{\rm Riess}$ neighborhood  enforces to pick too small values  $\Omo^0<0.27$ through extended contours that go beyond  $5\sigma$ c.l.} We also observe that the two ($w$)RVMs are much more compatible (already at $1\sigma$) with the $H_0^{\rm Planck}$ range than the $\CC$CDM. The latter, instead, requires some of the most external contours to reach the $H^{\rm Planck}_0$ $1\sigma$ region whether $H_0^{\rm Riess}$ is included or not in the fit. Thus, remarkably, in both cases when the full data string SNIa+BAO+$H(z)$+LSS+CMB enters the fit the $\CC$CDM has difficulties to overlap also with the $H_0^{\rm Planck}$ range at $1\sigma$, in contrast to the RVM and $w$RVM.
}
\end{center}
\end{figure}
\section{Discussion}
Following \cite{PRD2017}  the statistical analysis of the various models is performed in terms of a joint likelihood function, which is the product of the likelihoods for each data source
and includes the corresponding covariance matrices.
As indicated in the caption of Table 1,  the $\CC$CDM has $4$ parameters, whereas the XCDM and the DVMs have $5$, and finally any of the $w$DVMs has $6$. Thus, for a fairer comparison of the various nonstandard models with the concordance $\CC$CDM we have to invoke efficient criteria in which the presence of extra parameters in a given model is conveniently penalized so as to achieve a balanced comparison with the model having less parameters. The  Akaike information criterion (AIC) and the Bayesian information criterion (BIC) are known to be extremely valuable tools for a fair statistical analysis of this kind. They can be thought of as a modern quantitative formulation of Occam's razor. They read\,\cite{Akaike1974,Schwarz1978,KassRaftery1995}:
\begin{equation}\label{eq:AICandBIC}
{\rm AIC}=\chi^2_{\rm min}+\frac{2nN}{N-n-1}\,,\ \ \ \ \
{\rm BIC}=\chi^2_{\rm min}+n\,\ln N\,,
\end{equation}
where $n$ is the number of independent fitting parameters and $N$ the number of data points.
The bigger are the (positive) differences $\Delta$AIC and $\Delta$BIC with respect to the model having smaller values of AIC and BIC the higher is the evidence against the model with larger AIC and BIC. Take, for instance, Tables 1 and 2, where in all cases the less favored model is the $\CC$CDM (thus with larger AIC and BIC).
For $\Delta$AIC and $\Delta$BIC in the range $6-10$ one speaks of ``strong evidence'' against the $\CC$CDM, and hence in favor of the nonstandard models being considered. This is typically the situation for the RVM and $Q_{dm}$ vacuum models in Table 2 and for the three $w$DVMs in Table 1.
Neither the  XCDM nor the $Q_\CC$ vacuum model attain the ``strong evidence'' threshold in Tables 1 or 2. The XCDM parametrization, which is used as a baseline for comparison of the dynamical DE models, is nevertheless capable of detecting significant signs of dynamical DE, mainly in Table 1 (in which $H_0^{\rm Riess}$ is excluded), but not so in Table 2 (where $H_0^{\rm Riess}$ is included).  In contrast, model $Q_\CC$ does not change much from Table 1 to Table 2.

In actual fact, the vacuum model III ($Q_\CC$) tends to remain always fairly close to the $\CC$CDM. Its dynamics is weaker than that of the main DVMs (RVM and $Q_{dm}$). Being $|\nu_i|\ll 1$ for all the DVMs, the evolution of its vacuum energy density is approximately logarithmic: $\rL^{\rm III}\sim\rLo(1-3\nu_{\CC}\,\ln{a})$, as it follows from  (\ref{eq:rhowLdensities}) with $\epsilon=0$. Thus, it is significantly milder in comparison to that of the main DVMs, for which $\rL^{\rm I, II}\sim \rLo\left[1+(\Omega^{0}_m/\Omega^{0}_\Lambda)\nu_i (a^{-3}-1)\right]$. The performance of $Q_\CC$ can only be slightly better than that of $\CC$CDM, a fact that may have not been noted in previous studies -- see \,\cite{Salvatelli2014,Melchiorri2016,Murgia2016,Li2016,Melchiorri2017b} and references therein.

According to the same jargon, when the differences $\Delta$AIC and $\Delta$BIC are both above 10 one speaks of ``very strong evidence'' against the unfavored model (the $\CC$CDM, in this case), wherefore in favor of the dynamical vacuum and quasi-vacuum models. It is certainly the case of the RVM and $Q_{dm}$ models in Table 1, which are singled out as being much better than the $\CC$CDM in their ability to describe the overall observations. From Table 1 we can see that the best-fit values of $\nu_i$ for these models are secured at a confidence level of $\sim 3.8\sigma$. These two models are indeed the most conspicuous ones in our entire analysis, and remain strongly favored even if $H_0^{\rm Riess}$\,\cite{RiessH02016} is included (cf. Table 2). In the last case, the best-fit values of $\nu_i$ for the two models are still supported at a fairly large c.l. ($\sim 3.2\sigma$).  \newtext{This shows that the overall fit to the data in terms of dynamical vacuum is a real option since the fit quality is not exceedingly perturbed in the presence of the data point $H_0^{\rm Riess}$. However, the optimal situation is really attained in the absence of that point, not only because the fit quality is then higher but also because that point remains out of the fit range whenever the large scale structure formation data (LSS) are included. \newnewtext{For this reason we tend to treat that input as an outlier -- see also \cite{Lin2017} for an alternative support to this possibility, which we comment later on}. In the following, we will argue that a truly consistent picture with all the data is only possible for $H_0$ in the vicinity of  $H_0^{\rm Planck}$ rather than in that of $H_0^{\rm Riess}$.}

\newtext{The conclusion is that the $H_0^{\rm Riess}$-$H_0^{\rm Planck}$  tension cannot be relaxed without unduly forcing the overall fit, which is highly sensitive to the LSS data. It goes without saying that one cannot have a prediction that matches both $H_0$ regions at the same time, so at some point new observations (or the discovery of some systematic in one of the experiments) will help to consolidate one of the two ranges of values and exclude definitely the other. At present no favorable fit can be obtained from the $\CC$CDM that is compatible with any of the two $H_0$ ranges. This is transparent from Figs. 3 and 4, in which the $\CC$CDM remains always in between the two regions. However, our work shows that a solution (with minimum cost) is possible in terms of vacuum dynamics. Such solution, which inevitably puts aside the $H_0^{\rm Riess}$ range, is however compatible with all the remaining data and tends to favor the Planck range of $H_0$ values. The DVMs can indeed provide an excellent fit to the overall cosmological observations and be fully compatible with both the $H_0^{\rm Planck}$ value and at the same time with the needed low values of the $\sigma_8(0)$  observable, these low values of $\sigma_8(0)$ being crucial to fit the structure formation data. Such strategy is only possible in the presence of vacuum dynamics, whilst it is impossible with a rigid $\CC$-term, i.e. is not available to the $\CC$CDM.}


In Fig.\,1 we confront the various models with the LSS data when the local measurement $H_0^{\rm Riess}$ is not included in our fit. The differences can be better appraised in the plot on the right, where we observe that the RVM and $Q_{dm}$ curves stay significantly lower than the $\CC$CDM one (hence matching better the data than the $\CC$CDM), whereas those of XCDM and $Q_\CC$ remain in between.

Concerning the $w$DVMs, namely the quasi-vacuum models in which an extra parameter is at play (the EoS parameter $w$), we observe a significant difference as compared to the DVMs (with vacuum EoS $w=-1$): they {\it all} provide a similarly good fit quality, clearly superior to that of the $\CC$CDM (cf. Tables 1 and 2) but in all cases below that of the main DVMs (RVM and $Q_{dm}$), whose performance is outstanding.


\begin{figure}
\begin{center}
\label{contours}
\includegraphics[width=5in]{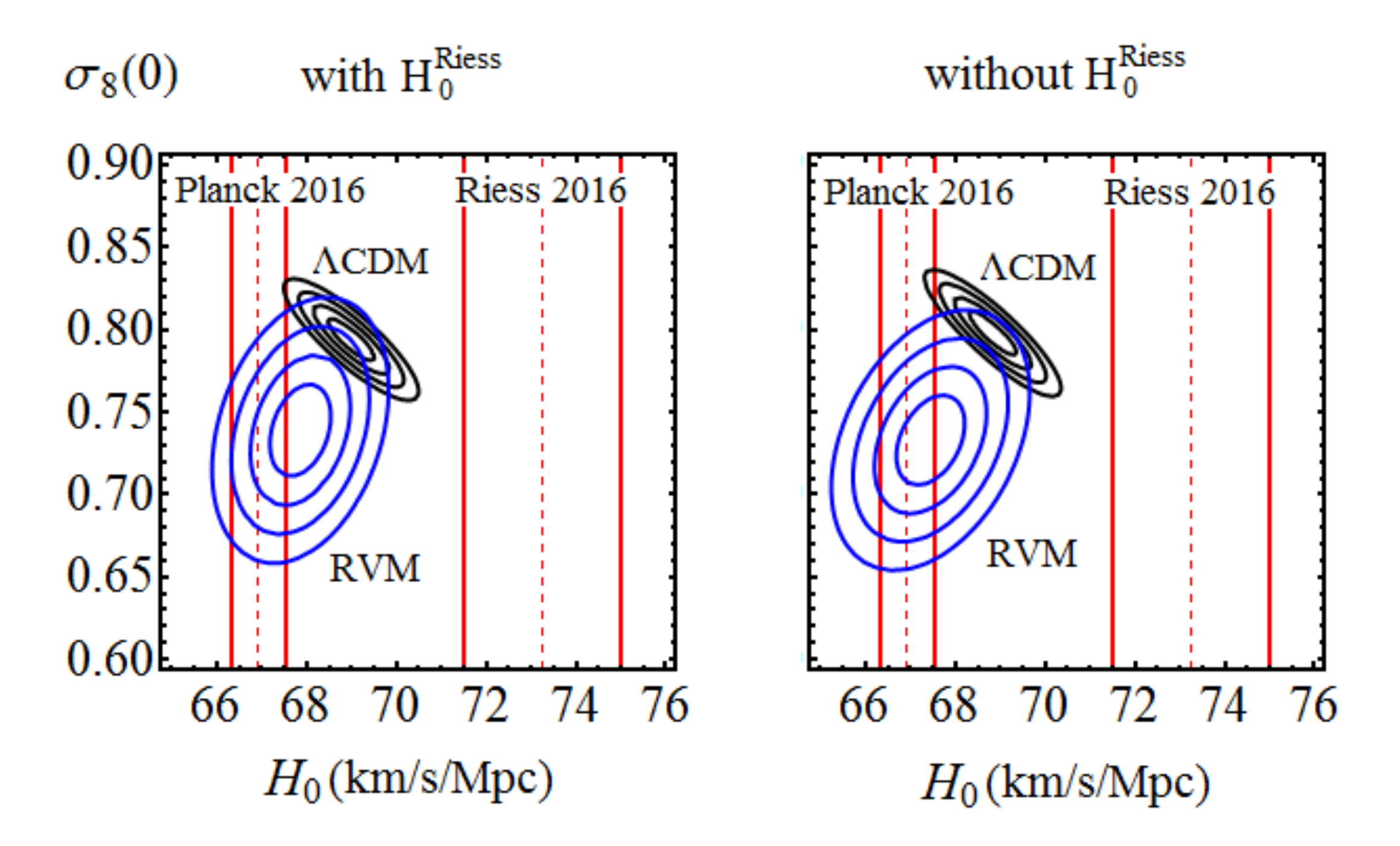}
\caption{\scriptsize \newtext{Contour lines for the $\CC$CDM (black) and RVM (blue) up to $4\sigma$ in the $(H_0,\sigma_8(0))$-plane. {As in Fig. 3, we present in the {\it left plot} the case when the local $H_0$ value of Riess {\it et al.}\,\cite{RiessH02016} is included in the fit (cf. Table 2), whereas in the {\it right plot} the case when that local value is {\it not} included (cf. Table 1). Again, any attempt at reaching the $H_0^{\rm Riess}$ neighborhood  enforces to extend the contours beyond the $5\sigma$ c.l.}, which would lead to a too low value of $\Omega_m^{0}$ in both cases (cf. Fig. 3) and, in addition, would result in a too large value of $\sigma_8(0)$ for the RVM. Notice that $H_0$ and $\sigma_8(0)$ are positively correlated in the RVM (i.e. $H_0$ decreases when $\sigma_8(0)$ decreases), whilst they are anticorrelated in the $\CC$CDM ($H_0$ increases when $\sigma_8(0)$ decreases, and vice versa). It is precisely this opposite correlation feature with respect to the $\CC$CDM what allows the RVM to improve the LSS fit in the region where both $H_0$ and $\sigma_8(0)$ are smaller than the respective $\CC$CDM values (cf. Fig. 1). This explains why the Planck range for $H_0$ is clearly preferred by the RVM, as it allows a much better description of the LSS data.}}
\end{center}
\end{figure}

In Table 3, in an attempt to draw our fit nearer and nearer to $H_0^{\rm Riess}$\,\cite{RiessH02016}, we test the effect of ignoring the LSS structure formation data, thus granting more freedom to the fit parameter space. We perform this test using the $\CC$CDM and models $(w)$RVM and $(w)Q_\CC$ (i.e. models I and III and testing both the vacuum and quasi-vacuum options), and we fit them to the CMB+BAO data alone. We can see that the fit values for $H_0$ increase in all starred scenarios (i.e. those involving the $H_0^{\rm Riess}$ data point in the fit), and specially for the cases Ia and IIIa in Table 3. Nonetheless, these lead to $\nu_i<0$ and $w<-1$ (and hence imply phantom-like DE); and, what is worse, the agreement with the LSS data is ruined (cf. Fig. 2) since the corresponding curves are shifted too high (beyond the $\CC$CDM one).  In the same figure we superimpose one more scenario, called IIIb, corresponding to a rather acute phantom behavior ($w=-1.184\pm0.064$). The latter was recently explored in \cite{Melchiorri2017b} so as to maximally relax the $H_0$ tension -- see also\,\cite{Melchiorri2016}. Unfortunately, we find (see Fig.\,2) that the associated LSS curve is completely strayed since it fails to minimally describe the $f\sigma_8$ data (LSS).

In Fig. 3 we demonstrate in a very visual way that, in the context of the overall observations (i.e.  SNIa+BAO+$H(z)$+LSS+CMB), whether including or not including the data point $H_0^{\rm Riess}$ (cf. Tables 1 and 2), it becomes impossible to getting closer to  the local measurement $H_0^{\rm Riess}$ unless we go beyond the $5\sigma$ contours and end up with a too low value $\Omo^0<0.27$. These results are aligned with those of \cite{Zhai2017}, in which the authors are also unable to accommodate the $H_0^{\rm Riess}$ value when a string of SNIa+BAO+$H(z)$+LSS+CMB data (similar but {\it not} equal to the one used by us) is taken into account. Moreover, we observe in Fig.\,3 not only that both the RVM and  $w$RVM remain much closer to  $H^{\rm Planck}_0$ than to $H_0^{\rm Riess}$, but also that they are overlapping with the $H^{\rm Planck}_0$  range much better than the $\CC$CDM does. {The latter is seen to have serious difficulties in reaching the Planck range unless we use the most external regions of the elongated contours shown in Fig.\,3.}

Many other works in the literature have studied the existing $H_0$ tension. For instance, in \cite{Wang2017} the authors find $H_0 = 69.13\pm 2.34$ km/s/Mpc assuming the $\Lambda$CDM model. {Such result almost coincides with the central values of $H_0$ that we obtain in Tables 1 and 2 for the $\Lambda$CDM.  This fact, added to the larger uncertainties of the result, seems to relax the tension. Let us, however, notice that the value of \cite{Wang2017} has been obtained using BAO data only, what explains the larger uncertainty that they find. In our case, we have considered a much more complete data set, which includes CMB and LSS data as well. This is what has allowed us to better constrain $H_0$ with smaller errors and conclude that when a larger data set (SNIa+BAO+$H(z)$+LSS+CMB)  is used, the fitted value of the Hubble parameter for the $\Lambda$CDM is incompatible with the Planck best-fit value at about $4\sigma$ c.l. Thus, the $\Lambda$CDM model seems to be in conflict not only with the local HST estimation of $H_0$, but also with the Planck one!}

\newtext{Finally, in Figs. 4 and 5 we consider the contour plots (up to $4\sigma$ and $3\sigma$, respectively) in the $(H_0,\sigma_8(0))$-plane for different situations. Specifically, in the case of Fig.\,4 the plots on the left and on the right are in exact correspondence with the situations previously presented in the left and right plots of Fig.\,3, respectively\footnote{The $H_0^{\rm Planck}$ band indicated in Figs. 3-5 is that of \cite{Planck2016}, which has no significant differences with that of \cite{Planck2015}.}. As expected, the contours in the left plot of Fig.\,4 are slightly shifted (``attracted'') to the right (i.e. towards the $H_0^{\rm Riess}$ region) as compared to those in the right plot because in the former $H_0^{\rm Riess}$ was included as a data point in the fit, whereas $H_0^{\rm Riess}$ was not included in the latter. Therefore, in the last case the contours for the RVM are more centered in the $H_0^{\rm Planck}$ region and at the same time centered at relatively low values of $\sigma_8(0)\simeq0.73-0.74$, which are precisely those needed for a perfect matching with the experimental data points on structure formation (cf. Fig. 1). On the other hand, in the case of Fig. 5 the contour lines correspond to the fitting sets Ia, IIIa of Table 3 (in which BAO and CMB data, but \emph{no} LSS formation data, are involved). As can be seen, the contour lines in Fig.\,5 can attain the Riess 2016 region for $H_0$, but they are centered at rather high values ($\sim 0.9$) of the parameter $\sigma_8(0)$. These are clearly higher than the needed values $\sigma_8(0)\simeq 0.73-0.74$. This fact demonstrates once more that such option leads to a bad description of the structure formation data.
The isolated point in Fig.\,5 is even worst: it corresponds to the aforementioned theoretical prediction for the scenario IIIb proposed in \cite{Melchiorri2017b}, in which the $H_0^{\rm Riess}$ region can be clearly attained but at the price of a serious disagreement with the LSS data. Here we can see, with pristine clarity, that such isolated point, despite it comfortably reaches the $H_0^{\rm Riess}$ region, it attains a value of  $\sigma_8(0)$ near $1$, thence completely strayed from the observations. This is, of course, the reason  why the upper curve in Fig.\,2 fails to describe essentially all points of the $f(z)\sigma_8(z)$ observable. So, as it turns, it is impossible to reach the $H_0^{\rm Riess}$ region without paying a high price, no matter what strategy is concocted to approach it in parameter space.}

\begin{figure*}[t!]
\begin{center}
\label{contours}
\includegraphics[width=3in]{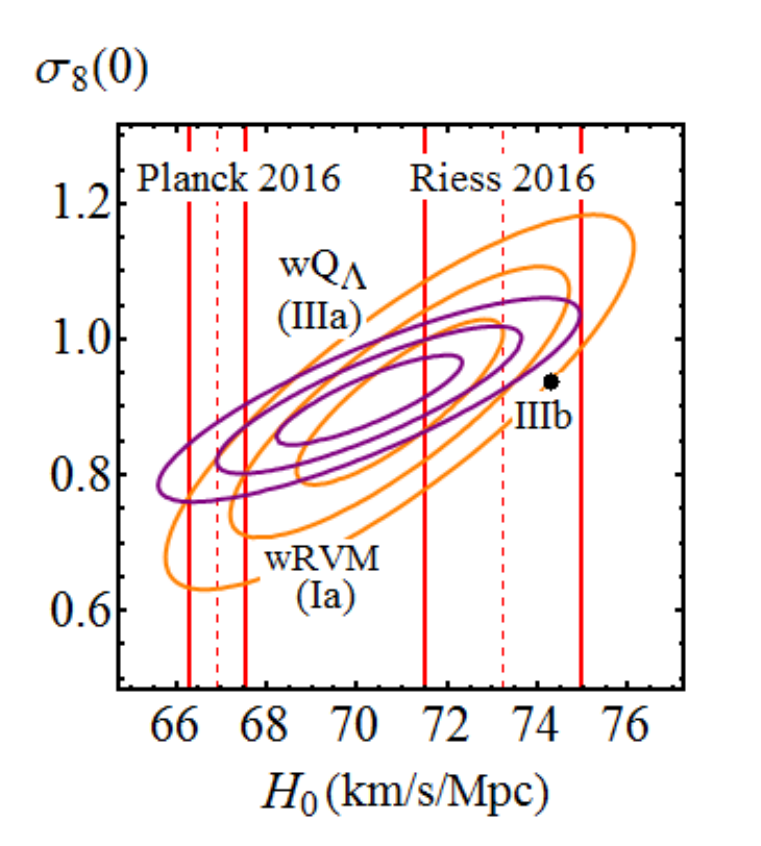}
\caption{\scriptsize \newtext{Contour lines for the models $w$RVM (Ia) and $w$Q$_\CC$ (IIIa) up to $3\sigma$ in the $(H_0,\sigma_8(0))$-plane, depicted in orange and purple, respectively, together with the isolated point (in black) extracted from the analysis of Ref. \cite{Melchiorri2017b}, which we call IIIb. The cases Ia, IIIa and IIIb correspond to special scenarios with $w\neq -1$ for Models I and III in which the value $H_0^{\rm Riess}$  is included as a data point and then a suitable strategy is followed to optimize the fit agreement with such value. The strategy consists to exploit the freedom in $w$ and remove the LSS data from the fit analysis. The plot clearly shows that some agreement is indeed possible, but only if  $w$ takes on values in the phantom region ($w<-1$) (see text) and at the expense of an anomalous (too large) value of the parameter $\sigma_8(0)$, what seriously spoils the concordance with the LSS data, as can be seen in Fig. 2.
}}
\end{center}
\end{figure*}

As indicated, we must still remain open to the possibility that the $H^{\rm Planck}_0$ and/or $H^{\rm Riess}_0$ measurements are affected by some kind of (unknown) systematic errors, although some of these possibilities may be on the way of  being ruled out by recent works. For instance, in \cite{Aylor2017} the authors study the systematic errors in Planck's data by comparing them with the South Pole Telescope data. Their conclusion is that there is no evidence of systematic errors in Planck's results. If confirmed, the class of the $(w)$RVMs studied here would offer a viable solution to both the $H_0$ and $\sigma_8(0)$ existing tensions in the data, which are both unaccountable within the $\CC$CDM. Another interesting result is the ``blinded''  determination of $H_0$  from \cite{Zhang2017}, based on a reanalysis of the SNIa and Cepheid variables data from the older work by Riess et al. \cite{RiessH02011}. These authors find $H_0 = 72.5\pm 3.2$ km/s/Mpc, which should be compared with $H_0 = 73.8\pm 2.4$ km/s/Mpc\,\cite{RiessH02011}. Obviously, the tension with $H^{\rm Planck}_0$ diminished since the central value decreased and in addition the uncertainty has grown by $\sim 33\%$. We should now wait for a similar reanalysis to be made on the original sample used in \cite{RiessH02016}, i.e. the one supporting the value  $H^{\rm Riess}_0$, as planned in \cite{Zhang2017}. \newnewtext{ In  \cite{Addison2017}  they show that by combining the latest BAO results with WMAP, Atacama Cosmology Telescope (ACT), or South Pole Telescope (SPT) CMB data produces values of $H_0$  that are $2.4-3.1\sigma$ lower than the distance ladder, independent of Planck. These authors conclude from their analysis  that it is not possible to explain the $H_0$  disagreement solely with a systematic error specific to the Planck data}. Let us mention other works, see e.g. \cite{Cardona2017,Feeney2017}, in which a value closer to $H_0^{\rm Riess}$ is found and the tension is not so severely loosened; or the work\,\cite{Follin2017}, which excludes systematic bias or uncertainty in the Cepheid calibration step of the distance ladder measurement by\,\cite{RiessH02016}. \newnewtext{Finally, we recall the aforementioned recent study \cite{Lin2017}, where the authors run a new (dis)cordance test to compare the constraints on $H_0$ from different methods and conclude that the local
measurement is an outlier compared to the others, what would favor a systematics-based explanation.}
Quite obviously, the search for a final solution to the $H_0$ tension is still work in progress.

\section{Conclusions}
\newnewtext{The present updated analysis of the cosmological data SNIa+BAO+$H(z)$+LSS+CMB disfavors the hypothesis $\CC=$const. as compared to the dynamical vacuum models (DVMs). This is consistent with our most recent studies\,\cite{ApJL2015,ApJ2017,MPLA2017,JSPRev2016,PRD2017}. Our results suggest a dynamical DE effect near $3\sigma$ within the standard XCDM parametrization and near $4\sigma$ for the best DVMs.  Here we have extended these studies in order to encompass the class of quasi-vacuum models ($w$DVMs), where the equation of state parameter $w$ is near (but not exactly equal) to $-1$. The new degree of freedom $w$ can then be used to try to further improve the overall fit to the data. But it can also be used to check if values of $w$ different from $-1$ can relax the existing tension between the two sets of  measurement of the $H_0$ parameter, namely those based:  i) on the CMB measurements by the Planck collaboration\,\cite{Planck2015,Planck2016}, and ii) on the local measurement (distance ladder method) using Cepheid variables\,\cite{RiessH02016}.}

\newnewtext{Our study shows that the RVM with $w=-1$ remains as the preferred DVM for the optimal fit of the data. At the same time it favors the CMB measurements of $H_0$ over the local measurement.  Remarkably, we find that not only the CMB and BAO data, but also the LSS formation data (i.e. the known data on $f(z)\sigma_8(z)$ at different redshifts), are essential to support the CMB measurements of $H_0$ over the local one. We have checked that if the LSS data are not considered (while the BAO and CMB are kept), then there is a unique chance to try to accommodate the local measurement of $H_0$, but only at the expense of a phantom-like behavior (i.e. for $w<-1$). In this region of the parameter space, however, we find that the agreement with the LSS formation data is manifestly lost, what suggests that the $w<-1$ option is ruled out. There is no other window in the parameter space where to accommodate the local $H_0$ value in our fit.  In contrast, when the LSS formation data are restored, the fit quality to the overall SNIa+BAO+$H(z)$+LSS+CMB observations improves dramatically and definitely favors the Planck range for $H_0$  as well as smaller values for $\sigma_8(0)$ as compared to the $\CC$CDM.}

\newnewtext{In short, our work suggests that signs of dynamical vacuum energy are encoded in the current cosmological observations. They  appear to be more in accordance with the lower values of $H_0$ obtained from the Planck (CMB) measurements than with the higher range of $H_0$ values obtained  from the present local (distance ladder) measurements, and provide smaller values of $\sigma_8(0)$ that are in better agreement with structure formation data as compared to the $\CC$CDM. We hope that with new and more accurate observations, as well as with more detailed analyses, it will be possible to assess the final impact of vacuum dynamics on the possible solution of the current tensions in the $\CC$CDM. }


\section{Acknowledgements}

We are partially supported by MINECO FPA2016-76005-C2-1-P, Consolider CSD2007-00042, 2014-SGR-104 (Generalitat de
Catalunya) and  MDM-2014-0369 (ICCUB).

\vspace{0.5cm}
{\bf Note Added:}
\newnewtext{Since the first version of this work appeared in preprint form,  arXiv:1705.06723, new analyses of the cosmological data have appeared, in particular the one-year results by the DES collaboration (DES Y1 for short)\,\cite{DES2017}. They do not find evidence for dynamical DE, and the Bayes factor indicates that the DES Y1 and Planck data sets are consistent with each other in the context of $\CC$CDM.
However, in our previous works -- see  in particular \,\cite{ApJ2017,PRD2017} -- we explained why the Planck results did not report evidence on dynamical DE. For instance, in \cite{Planck2015} they did not use  LSS (RSD) data, and in \cite{PlanckDE2015} they only used a limited set of  BAO and LSS points. In the mentioned works  \,\cite{ApJ2017,PRD2017} we have shown that under the same conditions we recover their results, but when we use the full data string, which involves not only CMB but also the rich  BAO+LSS data set, we do obtain instead positive indications of dynamical DE. A similar situation occurs with DES Y1; they do not use direct $f(z)\sigma_8(z)$  data on LSS structure formation despite they recognize that smaller values of $\sigma_8(0)$ than those predicted by the $\CC$CDM are necessary to solve the tension existing between the concordance model and the LSS observations. In contrast, let us finally mention that our positive result on dynamical DE is consistent with the recent analysis by Gong-Bo Zhao et al.\,\cite{GongBoZhao2017}, who report on a signal of dynamical DE at $3.5\sigma$ c.l using similar data ingredients as in our analysis.}

\end{document}